\documentclass[aps,twocolumn,floatfix,superscriptaddress]{revtex4}

\usepackage{amsmath}
\usepackage{amssymb}
\usepackage{nicefrac}
\usepackage{braket}
\usepackage[pdftex]{graphicx}
\usepackage{txfonts}
\usepackage{color}

\renewcommand{\vec}[1]{\boldsymbol{#1}}

\begin{document}

\title{Unconventional Topological Hall Effect in Skyrmion Crystals Caused by the Topology of the Lattice}

\author{B{\"o}rge G{\"o}bel}
\email[]{bgoebel@mpi-halle.mpg.de}
\affiliation{Max-Planck-Institut f\"ur Mikrostrukturphysik, D-06120 Halle (Saale), Germany}

\author{Alexander Mook}
\affiliation{Max-Planck-Institut f\"ur Mikrostrukturphysik, D-06120 Halle (Saale), Germany}

\author{J\"urgen Henk}
\affiliation{Institut f\"ur Physik, Martin-Luther-Universit\"at Halle-Wittenberg, D-06099 Halle (Saale), Germany}

\author{Ingrid Mertig}
\affiliation{Max-Planck-Institut f\"ur Mikrostrukturphysik, D-06120 Halle (Saale), Germany}
\affiliation{Institut f\"ur Physik, Martin-Luther-Universit\"at Halle-Wittenberg, D-06099 Halle (Saale), Germany}

\date{\today}

\begin{abstract}
The hallmark of a skyrmion crystal (SkX) is the topological Hall effect (THE). In this Article, we predict and explain an unconventional behavior of the topological Hall conductivity in SkXs. In simple terms, the spin texture of the skyrmions causes an inhomogeneous emergent magnetic field whose associated Lorentz force acts on the electrons. By making the emergent field homogeneous, the THE is mapped onto the quantum Hall effect (QHE). Consequently, each electronic band of the SkX is assigned to a Landau level. This correspondence of THE and QHE allows to explain the unconventional behavior of the THE of electrons in SkXs. For example, a skyrmion crystal on a triangular lattice exhibits a quantized topological Hall conductivity with steps of $2 \cdot e^2/h$ \emph{below} and with steps of $1 \cdot e^2/h$ \emph{above} the van Hove singularity. On top of this, the conductivity shows a prominent sign change \emph{at} the van Hove singularity. These unconventional features are deeply connected to the topology of the structural lattice.
\end{abstract}
    
\maketitle

\section{Introduction}
The quantum Hall effect (QHE) is one of the best known phenomena in condensed matter physics. It was first discussed for a two-dimensional electron gas in which the parabolic dispersion of free electrons is `compressed' into dispersionless Landau levels (LLs)~\cite{landau1930diamagnetismus,onsager1952interpretation}. Even before its experimental discovery~\cite{klitzing1980new} the QHE was described for various lattices in terms of Hofstadter butterflies~\cite{hofstadter1976energy,claro1979magnetic, rammal1985landau, claro1981spectrum, thouless1982quantized}: the quantized energy levels become dispersive and the Hall conductivity $\sigma_{xy}$ can change sign when applying a bias voltage. However, most of these manifestations of lattice topology remain to be verified by experiments. As an exception, $\sigma_{xy}$ of graphene has been measured in a small energy window for half filling~\cite{novoselov2005two}.  The observed unconventional quantization --- a sign change of the Hall conductivity for small variation of the bias --- has been understood in terms of Chern numbers of the LLs~\cite{hatsugai2006topological, sheng2006quantum}.

\begin{figure}
  \centering
  \includegraphics[width=1\columnwidth]{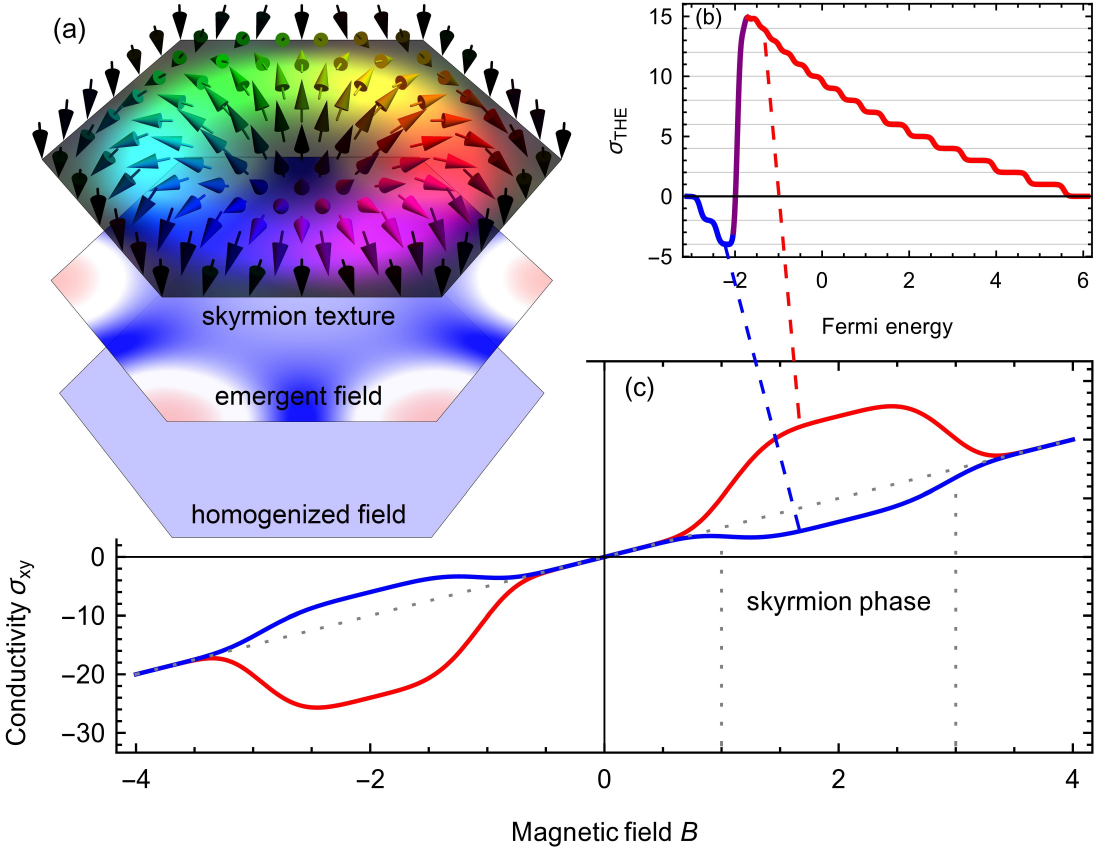}
  \caption{Core message of the Paper. (a) A skyrmion (top hexagon) generates an inhomogeneous emergent magnetic field (central hexagon; blue: positive; white: zero; red: negative). By redistributing this field such that it becomes homogeneous (lower hexagon), the topological Hall effect is mapped onto a quantum Hall effect. (b) Schematic bias dependence of the topological contribution $\sigma_\text{THE}$ to the Hall conductivity $\sigma_{xy}$, exhibiting a sign change at the energy of a van Hove singularity (purple). (c) Magnetic-field dependence of $\sigma_{xy}$ for a bias below (blue) and above (red) the van Hove singularity. $\sigma_\text{xy}$ can show a decrease (blue) as well as an increase (red) in the skyrmion crystal phase which is present for $1 \le B \le 3 $ in arbitrary units.}
  \label{fig:skyrmion_cell}
\end{figure}

Skyrmions~\cite{skyrme1962unified} have conquered the field of magnetism since their theoretical~\cite{bogdanov1989thermodynamically, bogdanov1994thermodynamically, rossler2006spontaneous} and experimental discoveries~\cite{muhlbauer2009skyrmion}. They are typically generated by the Dzyaloshinskii-Moriya interaction~\cite{dzyaloshinsky1958thermodynamic,moriya1960anisotropic} in chiral magnets, for example in the non-centrosymmetric B20 materials, prominently represented by MnSi~\cite{muhlbauer2009skyrmion}. A skyrmion spin texture $\vec{s}(\vec{r})$ [arrows in Fig.~\ref{fig:skyrmion_cell}(a)] stands out from topologically trivial textures (e.\,g., collinear magnets or spin helices) by its topological charge
\begin{align*}
	N_\mathrm{Sk} = \frac{1}{4\pi} \int_{xy} n_\mathrm{Sk}(\vec{r})\, \mathrm{d}^{2} r,
	\quad 
	n_\mathrm{Sk} (\vec{r}) = \vec{s}(\vec{r}) \cdot \left[ \frac{\partial \vec{s}(\vec{r})}{\partial x}  \times  \frac{\partial \vec{s}(\vec{r})}{\partial y}  \right],
\end{align*}
which is a nonzero integer; $n_\mathrm{Sk}(\vec{r})$ is the topological charge density.  It gives rise to the topological Hall effect (THE)~\cite{neubauer2009topological,schulz2012emergent,kanazawa2011large, lee2009unusual,li2013robust,hamamoto2015quantized,lado2015quantum}: the nontrivial magnetic texture causes an emergent magnetic field $\vec{B}_\mathrm{em}$ which acts on the propagating electrons by  its Lorentz force.

In this Article, we discuss an unconventional quantized THE in skyrmion crystals (SkXs): the topological Hall conductivity exhibits a prominent change of sign as a function of bias voltage. This sign change is assigned to the \emph{topology of the structural lattice} rather than to the nontrivial magnetic topology of the skyrmions. For this purpose, we relate the THE to the QHE as follows. The emergent field $\vec{B}_\mathrm{em}$ is inhomogeneous [central hexagon in Fig.~\ref{fig:skyrmion_cell}(a)] because it is proportional to $n_\mathrm{Sk}(\vec{r})$. By making  $\vec{B}_\mathrm{em}$ homogeneous, i.\,e., by redistributing the magnetic flux, the THE is mapped onto a QHE on a structural lattice but in an homogeneous field [lower hexagon in Fig.~\ref{fig:skyrmion_cell}(a)]. As a result, the quantum Hall conductivity $\sigma_{xy}$ is quantized in the same way as for the THE: in steps of $2 \cdot e^2/h$ below and in steps of $1 \cdot e^2/h$ above the van Hove singularity (VHS). Most strikingly, it exhibits an abrupt change of sign when the VHS is crossed in dependence of bias. We attribute this unconventional behavior to the number and the character (electron versus hole) of the Fermi pockets. Thus, it is tightly related to the topology of the structural lattice. 

The exceptional behavior, sketched in Fig.~\ref{fig:skyrmion_cell}(b), calls for experiments on samples exhibiting a SkX phase. The Hall conductivity $\sigma_{xy}$ in clean samples (mean free path of the electrons is larger than the skyrmion size) should be extremely sensitive to a gate voltage: the contribution of the THE to $\sigma_{xy}$ can change sign [red versus blue curve in Fig.~\ref{fig:skyrmion_cell}(c)]. In the following, we provide details supporting our claim.

\section{Electrons in a skyrmion crystal}
Following Ref.~\onlinecite{hamamoto2015quantized}, we describe the spin-dependent electronic structure by means of the tight-binding Hamiltonian
\begin{align} 
  H & = \sum_{ij} t_{ij} \,c_{i}^\dagger c_{j} + m \sum_{i} \vec{s}_{i} \cdot (c_{i}^\dagger \vec{\sigma}c_{i})
  \label{eq:ham_the} 
\end{align}
with constant nearest-neighbor hopping $t_{ij} = t$ ($i$, $j$ sites of the structural lattice). The electron spin is coupled to the skyrmion magnetic texture $\{ \vec{s}_{i} \}$ with strength $m$ (measured in units of $t$; second sum). $\vec{\sigma}$ is the vector of Pauli matrices, $c_{i}^\dagger$ and $c_{i}$ are spin-dependent creation and annihilation operators, respectively. 

To model a SkX (a regular array of skyrmions), $\{ \vec{s}_{i} \}$ is assumed to be a triple-$q$ state~\cite{okubo2012multiple}, that is, a coherent superposition of three spin spirals with a prescribed wavelength $\lambda$. In the following, we consider a structural \emph{triangular lattice} with lattice constant $a$. 

The intrinsic contribution to the Hall conductivity~\cite{nagaosa2010anomalous}
\begin{align}
  \sigma_{xy}(E_\mathrm{F}) & = \frac{e^{2}}{h} \frac{1}{2\pi} \sum_{n} \int_{\mathrm{BZ}} \Omega_{n}^{(z)}(\vec{q}) \, f(E_{n}(\vec{q}) - E_\mathrm{F}) \,\mathrm{d}^{2}q \label{eq:cond}
\end{align}
is given by a Brillouin-zone (BZ) integral of the Berry curvature $\Omega_{n}^{(z)}(\vec{q}) = \partial_{q_{x}} A_{n}^{(y)}(\vec{q}) - \partial_{q_{y}}A_{n}^{(x)}(\vec{q})$. The Berry connection $\vec{A}_{n}(\vec{q}) = - \mathrm{i} \braket{u_{n}(\vec{q})| \nabla_{\vec{q}} | u_{n}(\vec{q})}$ is determined from the eigenvectors $u_{n}(\vec{q})$ with energies $E_{n}(\vec{q})$ of the Hamiltonian~\eqref{eq:ham_the}. The sum runs over all bands $n$;  $e$ and $h$ are the electron charge and Planck's constant, respectively, while $f(E)$ is the Fermi distribution function.

At zero temperature only states below the Fermi energy $E_\mathrm{F}$ contribute to the transport: if $E_\mathrm{F}$ is located within the band gap above the $l$-th band, $\sigma_{xy}$ is proportional to the winding number $w_{l} = \sum_{n \le l} C_n$ (Refs.~\onlinecite{Hatsugai1993,Hatsugai1993a}) which is the accumulation of the Chern numbers 
\begin{align}
	C_{n} = \frac{1}{2\pi} \int_{\mathrm{BZ}}\Omega_{n}^{(z)}(\vec{q})\, \mathrm{d}^{2}q.
\end{align}

\section{Quantized topological Hall effect}
For zero coupling [$m = 0$  in eq.~\eqref{eq:ham_the}], the bands are spin-degenerate and we obtain the band structure of a triangular lattice. The bottom of the band is at $-3 t$,  its top at $+6 t$;  a VHS shows up at  $E_\mathrm{VHS} \equiv -2 t$ [Fig.~\ref{fig:QHE_series}(b)].

\begin{figure}
  \centering
  \includegraphics[width=1\columnwidth]{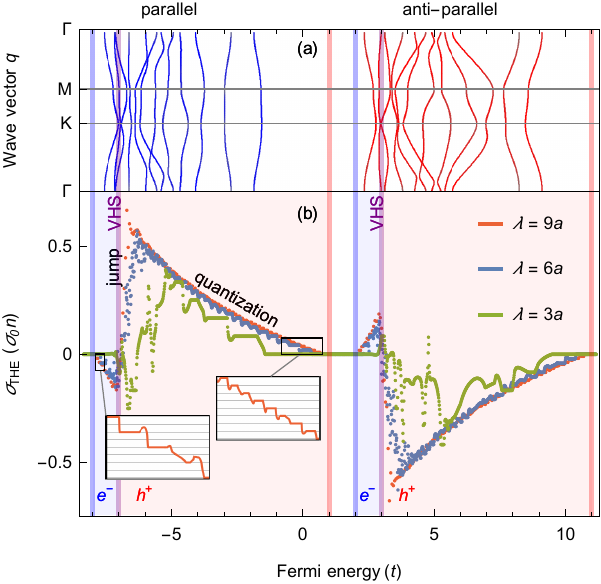}
  \caption{Topological Hall effect in SkXs. (a) Electronic band structure of a SkX with skyrmion size $\lambda = 3 a$ for coupling strength $m = 5 t$ ($a$ lattice constant). The alignment of the electron spin to the magnetic skyrmion texture is indicated by color (parallel: blue; antiparallel: red). (b) Topological Hall conductivity $\sigma_\text{THE}$ versus Fermi energy for skyrmion sizes $\lambda = 3 a$ (green), $6 a$ (blue), and $9 a$ (red). $\sigma_\text{THE}$ is normalized to the number $n$ of atomic sites per skyrmion unit cell ($\sigma_{0} = e^2 / h$ conductance quantum). Energy regions with electron- ($e^{-}$, blue background) and hole-like ($h^{+}$, red background) behavior are indicated (see text).}
\label{fig:pic_THE}
\end{figure}

For finite coupling $m$, the electron spin aligns with the skyrmion spin texture and the spin degeneracy is lifted. With increasing $m$, the band structure is split into two blocks of bands: one with spins parallel, the other with spins antiparallel to the spin texture. In the limit $m \rightarrow \infty$ both blocks exhibit identical dispersion relations.

The band structure for  $m = 5 t$ and $\lambda = 3 a$ is depicted in Fig.~\ref{fig:pic_THE}(a). In each block, the energetically higher bands are well separated and show considerable dispersion (right part of each block). Close to the VHSs, that is at $E = -2 t \pm m$, the bands become very narrow. 

The separation into blocks is reflected in the conductivity $\sigma_{xy}$ [Fig.~\ref{fig:pic_THE}(b)]. Since both blocks produce similar features, except for a change of sign, it is sufficient to discuss the block with lower energy. Starting from the band bottom, $\sigma_{xy}$ is negative and decreases with energy in quanta of $2 \cdot e^2 / h$. Close to the VHS ($E \approx -2 t - m = -7 t$), the conductivity increases abruptly to positive values. At larger energies $\sigma_{xy}$ drops again but in steps of $1 \cdot e^2 / h$ until it reaches zero conductance. This `quantization' region [label in Fig.~\ref{fig:pic_THE}(b)] shows up most pronounced for small skyrmions; cf.\ $\lambda = 3 a$ (green data set). Recall that there  the bands are well separated by gaps, the associated states carry Chern number $-1$. 

The saw-tooth shape of $\sigma_{xy}$ becomes more pronounced the larger the skyrmion size $\lambda$: the steps as well as the jump become energetically more narrow because there are more bands within the same energy range [compare the green, blue, and red data sets in Fig.~\ref{fig:pic_THE}(b)].

\subsection{Transformation to the emergent field}
In the strong-coupling limit $m \gg t$, the electron spin is fully aligned with the skyrmion texture and the two blocks of bands are identical but rigidly shifted in energy by the Zeeman term in the Hamiltonian. Consequently, each individual block can be discussed in terms of \emph{spinless} electrons. However, the skyrmion texture has to be taken into account by a local gauge transformation into the reference frame of its magnetic moments (Refs.~\onlinecite{everschor2014real, hamamoto2015quantized, ohgushi2000spin}). The gauge field $\vec{A}(\vec{r})$ defines the emergent magnetic field $\vec{B}_{\mathrm{em}}(\vec{r}) = \vec{\nabla} \times \vec{A}(\vec{r})$ with $B_{\mathrm{em}}^{(z)}(\vec{r}) \propto n_\mathrm{Sk} (\vec{r})$ \cite{everschor2014real}, which is collinear (along $z$) but inhomogeneous [central hexagon in Fig.~\ref{fig:skyrmion_cell}(a)].  The gauge transformation recasts the coupling of the electron spin to the skyrmion texture to a fictitious field acting on the electron charge.  Of course, both descriptions yield identical results for the THE\@. 

In the tight-binding model, the gauge field $\vec{A}(\vec{r})$ leads to effective complex hopping strengths~\cite{hamamoto2015quantized}
\begin{align}
  t^\mathrm{eff}_{ij} & \equiv t \cos{\frac{\theta_{ij}}{2}} \mathrm{e}^{\mathrm{i} a_{ij}} \label{eq:discreteaij}
\end{align}
that enter the Hamiltonian of the quantum Hall effect
\begin{align}
	H_\text{QH}  = \sum_{ij}t^\mathrm{eff}_{ij} \,d_{i}^\dagger d_{j}. \label{eq:QHE-Ham}
\end{align}
$d_{i}^\dagger$ ($d_{i}$) is  a creation (annihilation) operator and $\theta_{ij}$ the angle between the spins at sites $i$ and $j$. With the polar angles $\phi_{i}$ and $\phi_{j}$ of these spins the phase in eq.~\eqref{eq:discreteaij} is written as~\cite{hamamoto2015quantized} 
\begin{align}
	\tan a_{ij} = - \frac{\sin(\phi_{i}-\phi_{j})}{\cos(\phi_{i}-\phi_{j})+\cot\frac{\theta_{i}}{2}\cot\frac{\theta_{j}}{2}}. \label{eq:aij}
\end{align}

\subsection{Topological Hall effect as quantum Hall effect}
The reformulation of the THE as QHE requires to redistribute the inhomogeneous emergent field $B_\mathrm{em}^{(z)}(\vec{r})$  into a \emph{homogeneous} field with strength $B$. The topological charge of each skyrmion is conserved by the constraint $(2 \pi)^{-1} \int_\mathrm{uc} B_\mathrm{em}^{(z)} \,\mathrm{d}^{2}r = 1$ (uc unit cell of the SkX). The hopping strengths  in Eq.~\eqref{eq:discreteaij} and especially the phases  $a_{ij}$ [Eq.~\eqref{eq:aij}] have to be adjusted accordingly~\cite{hamamoto2015quantized},
\begin{align}
  t^\mathrm{eff}_{ij} & = t\,\exp\left( -\mathrm{i} e / \hbar \int_{\vec{r}_i}^{\vec{r}_j} \vec{A}(\vec{r}) \cdot \mathrm{d}\vec{l} \right).
\end{align}
$\text{d}\vec{l}$ points along the hopping path ($\vec{r}_i \to \vec{r}_j$) and $\vec{A}$ is the vector potential of the homogeneous magnetic field with $\vec{B} = \vec{\nabla} \times \vec{A}$. For our calculations we used $\vec{A}(\vec{r})=B\,\vec{e}_x (y-x/\sqrt{3})$.

\begin{figure}
  \centering
  \includegraphics[width=1\columnwidth]{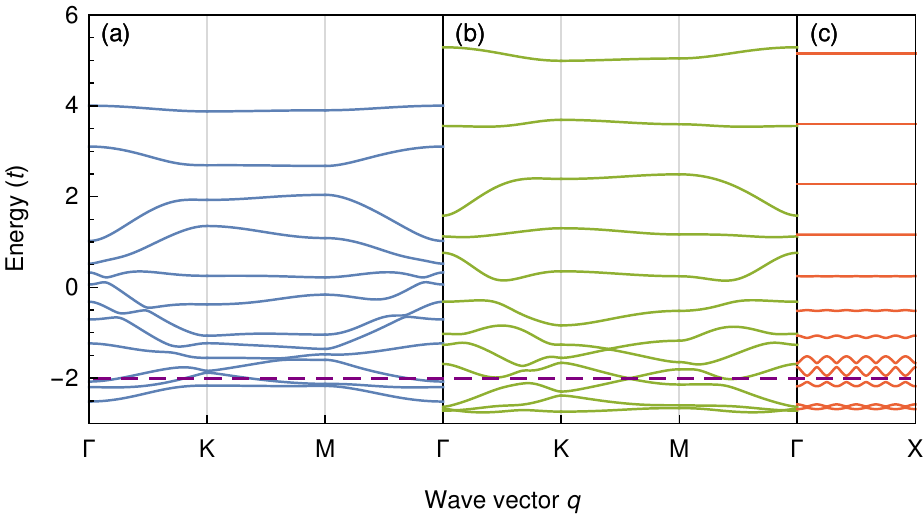}
  \caption{Electronic band structure of a skyrmion crystal and Landau levels. (a) Band structure of a skyrmion crystal (skyrmion size $\lambda = 3 a$, 12 sites per unit cell). (b) As (a) but with the approximation $\cos\theta_{ij}/2 \rightarrow 1$ in eq.~\eqref{eq:discreteaij}. (c) Landau levels for a homogeneous emergent magnetic field. The five topmost bands in (a)--(c) carry Chern number $-1$. The energy of the van Hove singularity is indicated by the purple dashed line.}
  \label{fig:pic_qhe}
\end{figure}

It is illustrative to compare the band structures for the inhomogeneous and the homogeneous emergent field (Fig.~\ref{fig:pic_qhe}). The total band width for the inhomogeneous emergent field (a) is increased if the term $\cos\theta_{ij} / 2$ in eq.~\eqref{eq:discreteaij} is approximated by $1$ (b); however, the shapes of the individual bands remain almost unchanged. The total band width in (b) is very close to that of the LLs (c). On top of that,  there is a one-to-one correspondence between the bands in (b) and the LLs (c). This is most obvious for large energies where the Chern numbers ($-1$) are identical as well. Hence, we conclude that the redistribution of the emergent field merely causes band width broadening but conserves the topology. 

\begin{figure*}
  \centering
  \includegraphics[width=2\columnwidth]{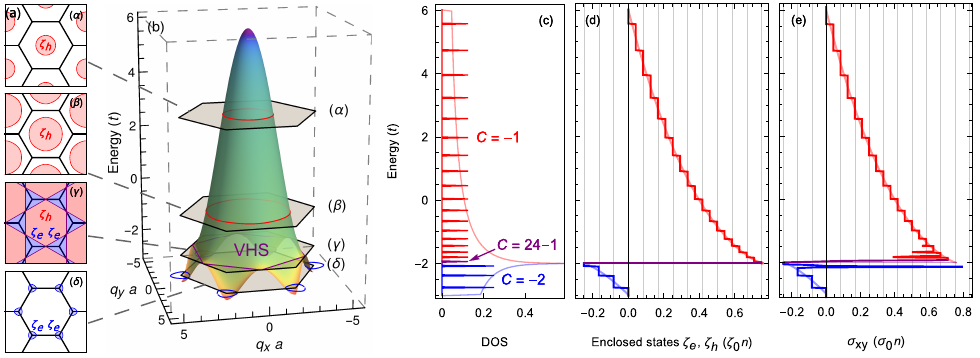}
  \caption{Quantum Hall effect on the triangular lattice for $n=24$ sites in the unit cell. (a) and (b) The band structure for the triangular lattice without magnetic field is depicted in (b). Cuts at selected energies are labeled $(\alpha), \ldots, (\delta)$ and are shown in (a). The numbers $\zeta_{\mathrm{e}}$ and $\zeta_{\mathrm{h}}$ of enclosed states  in the Brillouin zone (black hexagons) has to obey Onsager's quantization scheme. At the van Hove singularity [$E_\mathrm{VHS} = -2  t$, cut $(\gamma)$] the constant-energy contours exhibit a Lifshitz transition. (c) Density of states (DOS) of the band structure in (b) depicted by light smooth curves. The associated Landau levels are shown by dark colors;  their Chern numbers $C$ are indicated. (d) Number of enclosed states $\zeta_{\mathrm{e}}$ and $\zeta_{\mathrm{h}}$  before (light smooth curves) and after Landau quantization (dark). (e) Transverse quantum Hall conductivity in units of $\sigma_{0} = e^2 / h$. In all panels, the character of the constant-energy pockets is indicated by color (blue: electron-like; red: hole-like).}
  \label{fig:QHE_series}
\end{figure*}

We now corroborate the close relation of  THE and QHE further. Constant-energy cuts (CECs) through the original band structure of the triangular lattice [$(\alpha), \ldots, (\delta)$ in Fig.~\ref{fig:QHE_series}(a) and~(b)] at elevated energies are circular because there the band is parabolic [cf.\ CEC $(\alpha)$]. Separating occupied states in the outside from unoccupied states in the inside of the CEC, a circle is a hole pocket with negative curvature (for $t > 0$). The LLs in this energy region are dispersionless, as expected for free electrons.  

Constant energy cuts closer to the VHS show increased hexagonal warping [cf.\ CEC $(\beta)$]. At the VHS the CEC is a hexagon [cf.\ CEC $(\gamma)$]. Having no curvature, $(\gamma)$ features an infinite effective mass, with the consequence that electrons at the VHS are not affected by the emergent field. These electrons behave as in a continuum~\cite{hsu1976level}; the associated band exhibits oscillations in the reduced zone scheme [Fig.~\ref{fig:pic_qhe}(c)], thereby resembling the band structure of the triangular lattice [Fig.~\ref{fig:QHE_series}(b)]. Constant-energy cuts at energies below the VHS exhibit two electron orbits [cf.\ CEC $(\delta)$].

Landau levels with energies larger than that of the VHS carry Chern number $-1$ [red peaks in Fig.~\ref{fig:QHE_series}(c)] because there is a single hole-like Fermi contour [$(\alpha)$ and $(\beta)$ in Fig.~\ref{fig:QHE_series}(a); the number of enclosed states is $\zeta_{\mathrm{h}}$]. In contrast, Landau levels below the VHS appear in pairs because there are two electron-like Fermi lines, each enclosing $\zeta_{\mathrm{e}}$ states [see $(\delta)$]. Thus, each pair carries twice the Chern number of free-electron-like LLs, i.\,e. $-2$ (blue peaks).

To explore the Hall conductivity in detail, we utilize an approximate \emph{construction}. Onsager's quantization scheme~\cite{onsager1952interpretation} allows to deduce LLs directly from the CECs of the original band structure of the triangular lattice [Fig.~\ref{fig:QHE_series}(b)]: if a LL contains $\zeta_{0}$ states, the associated CEC encloses $(j + \nicefrac{1}{2}) \zeta_{0}$ states ($j$ integer). This means for CECs with two electron orbits that the total number of enclosed states reads $2 \cdot (j + \nicefrac{1}{2}) \zeta_{0}$. The character of the pockets is respected by assigning positive numbers to $\zeta_{\mathrm{h}}$ for hole-like pockets (red in Fig.~\ref{fig:QHE_series}) but negative numbers to $\zeta_{\mathrm{e}}$ for electron-like pockets (blue in Fig.~\ref{fig:QHE_series})~\cite{arai2009quantum,lifshitz1957theory}. At the van Hove singularity which separates hole- from electron-like pockets the constructed conductivity changes sign [Fig.~\ref{fig:QHE_series}(d)].

The conductivity constructed from the number of enclosed states is remarkably similar to the quantum Hall conductivity $\sigma_{xy}$ that has been explicitly computed from eq.~\eqref{eq:cond} for the LLs [Figs.~\ref{fig:QHE_series}(d) versus (e)]. The Chern numbers are proportional to the number of pockets and appear as steps in the constructed curve, although they are never explicitly used for the construction. Both curves deviate near the VHS at $E_\mathrm{VHS} = -2 t$, that is, where lattice effects are prominent; recall that the latter are neglected in Onsager's quantization scheme. The Landau levels in this energy range  show oscillations  [Fig.~\ref{fig:pic_qhe}(c)]. Thus, the sign change of $\sigma_{xy}$ is not located exactly at the VHS but is associated with the oscillating LL closest to the VHS\@. 

This particular LL is composed of states with energies below \emph{and} above the VHS; corresponding constant-energy cuts are taken below and above the cut $(\gamma)$ [Figs.~\ref{fig:QHE_series}(a) and (b)], which features an open orbit with infinite mass. The distinction of the number of enclosed states and of their character---two hole-like pockets below the VHS but a single electron-like pocket above the VHS---dictates a mismatch of Chern numbers. The result is a large Chern number of $C = n - 1$ for this particular LL [magenta peak in Fig.~\ref{fig:QHE_series}(c) with $C = +23$ for $n=24$] \cite{hatsugai2006topological,sheng2006quantum, arai2009quantum,arai2010numerical}. The outstanding Chern number compensates the sum of all other Chern numbers. This result is clearly a manifestation of the van Hove singularity. It is thus caused by the topology of the structural lattice: the large Chern number and the associated jump of the transverse Hall conductivity would occur at VHSs for \emph{every} lattice.

The energy dependence of the quantum Hall conductivity $\sigma_{xy}$ shows striking similarity to that of the topological Hall effect in skyrmion crystals. Both conductivities---$\sigma_{xy}$ of one block in Fig.~\ref{fig:pic_THE}(b) and $\sigma_{xy}$ in Fig.~\ref{fig:QHE_series}(e)---feature steps of $-2 \cdot e^2 / h$ below the VHS, the substantial jump near the VHS, and steps of $-1 \cdot e^2 / h$ above the VHS\@.  Accordingly, topological and quantum Hall effect are essentially equivalent.  A difference is that in the case of the THE the inhomogeneity of the emergent field `adds' dispersion to the bands [cf. Fig.~\ref{fig:pic_qhe}(b) and (c)]. 
To reiterate, the effects ascribed to the topology of the structural lattice---quantization and the jump of $\sigma_{xy}$--- are transferred from the THE to the QHE and \textit{vice versa}. In general, topological Hall conductivities would rise abruptly at VHSs on every lattice.

\section{Suggestion for experiments}
The established relation of  lattice topology and bias dependence of the topological Hall conductivity calls for experimental verification. 
The quantized topological Hall effect can be studied in metals which exhibit a SkX phase, e.\,g., MnSi (Ref.~\onlinecite{muhlbauer2009skyrmion}), Fe$_{1-x}$Co$_{x}$Si (Ref.~\onlinecite{yu2010real}), and FeGe (Ref.~\onlinecite{yu2011near}). A necessary prerequisite is that the mean free path of the electrons is larger than the skyrmion size.

In samples with insignificant anomalous Hall effect, the Hall conductivity $\sigma_{xy}$ increases with $B$, if $B$ is small. A transition from a topologically trivial magnetic phase to a SkX phase would cause an abrupt change of $\sigma_{xy}$ because the THE provides an additional contribution to $\sigma_{xy}$ [Fig.~\ref{fig:skyrmion_cell}(b)].  The application of a gate voltage, which allows to scan the energy dependence of $\sigma_{xy}$, can make this variation either a decrease or an increase, depending on whether the Fermi energy lies below or above a VHS [red and blue lines in Fig.~\ref{fig:skyrmion_cell}(c)]. 

The change of sign in $\sigma_{xy}$ is preferably studied for large skyrmions: the sawtooth-shaped variation of the Hall conductivity becomes cultrate with increasing skyrmion size [green, blue, and red curves in Fig.~\ref{fig:pic_THE}(b)]. This behavior is, however, limited by the finite sample size because an experiment measures a conductance rather than a conductivity. Therefore, a compromise between signal strength (favored by small skyrmions) and sharpness of the sawtooth-shaped feature (favored by large skyrmions) has to be made.

In real samples, the Hall conductivity is due to two contributions: the topological Hall effect and the anomalous Hall effect. The THE contribution to the Hall conductivity is attributed to the nontrivial topology in reciprocal space that arises from the real-space topology of the magnetic texture; spin-orbit coupling is not required. The contribution of the anomalous Hall effect relies on a nonzero Berry curvature as well but is solely induced by intrinsic spin-orbit coupling and a topologically trivial magnetic texture (like a ferromagnet); a topologically nontrivial magnetic texture is not required. Both anomalous and topological contributions to the Hall conductivity would vary with gate voltage. Thus, the dominating contribution of the two effects should be identified in advance~\cite{matsuno2016interface}. As real materials exhibit complicated band structures and feature intrinsic spin-orbit interaction, a combined analysis of the anomalous and the topological Hall effects seems to be worthwhile in the future.

\begin{acknowledgments}
Fruitful discussions with Naoto Nagaosa and Tom{\'a}{\v{s}} Rauch are gratefully acknowledged. This work is supported by SPP 1666 of Deutsche Forschungsgemeinschaft (DFG).
\end{acknowledgments}

\bibliography{short,MyLibrary}

\begin{thebibliography}{38}
\expandafter\ifx\csname natexlab\endcsname\relax\def\natexlab#1{#1}\fi
\expandafter\ifx\csname bibnamefont\endcsname\relax
  \def\bibnamefont#1{#1}\fi
\expandafter\ifx\csname bibfnamefont\endcsname\relax
  \def\bibfnamefont#1{#1}\fi
\expandafter\ifx\csname citenamefont\endcsname\relax
  \def\citenamefont#1{#1}\fi
\expandafter\ifx\csname url\endcsname\relax
  \def\url#1{\texttt{#1}}\fi
\expandafter\ifx\csname urlprefix\endcsname\relax\def\urlprefix{URL }\fi
\providecommand{\bibinfo}[2]{#2}
\providecommand{\eprint}[2][]{\url{#2}}

\bibitem[{\citenamefont{Landau}(1930)}]{landau1930diamagnetismus}
\bibinfo{author}{\bibfnamefont{L.}~\bibnamefont{Landau}}, \bibinfo{journal}{Z.
  Phys.} \textbf{\bibinfo{volume}{64}}, \bibinfo{pages}{629}
  (\bibinfo{year}{1930}).

\bibitem[{\citenamefont{Onsager}(1952)}]{onsager1952interpretation}
\bibinfo{author}{\bibfnamefont{L.}~\bibnamefont{Onsager}},
  \bibinfo{journal}{The London, Edinburgh, and Dublin Philosophical Magazine
  and Journal of Science} \textbf{\bibinfo{volume}{43}}, \bibinfo{pages}{1006}
  (\bibinfo{year}{1952}).

\bibitem[{\citenamefont{Klitzing et~al.}(1980)\citenamefont{Klitzing, Dorda,
  and Pepper}}]{klitzing1980new}
\bibinfo{author}{\bibfnamefont{K.~v.} \bibnamefont{Klitzing}},
  \bibinfo{author}{\bibfnamefont{G.}~\bibnamefont{Dorda}}, \bibnamefont{and}
  \bibinfo{author}{\bibfnamefont{M.}~\bibnamefont{Pepper}},
  \bibinfo{journal}{Phys.\ Rev.\ Lett.} \textbf{\bibinfo{volume}{45}},
  \bibinfo{pages}{494} (\bibinfo{year}{1980}).

\bibitem[{\citenamefont{Hofstadter}(1976)}]{hofstadter1976energy}
\bibinfo{author}{\bibfnamefont{D.~R.} \bibnamefont{Hofstadter}},
  \bibinfo{journal}{Phys.\ Rev.\ B} \textbf{\bibinfo{volume}{14}},
  \bibinfo{pages}{2239} (\bibinfo{year}{1976}).

\bibitem[{\citenamefont{Claro and Wannier}(1979)}]{claro1979magnetic}
\bibinfo{author}{\bibfnamefont{F.}~\bibnamefont{Claro}} \bibnamefont{and}
  \bibinfo{author}{\bibfnamefont{G.}~\bibnamefont{Wannier}},
  \bibinfo{journal}{Phys.\ Rev.\ B} \textbf{\bibinfo{volume}{19}},
  \bibinfo{pages}{6068} (\bibinfo{year}{1979}).

\bibitem[{\citenamefont{Rammal}(1985)}]{rammal1985landau}
\bibinfo{author}{\bibfnamefont{R.}~\bibnamefont{Rammal}},
  \bibinfo{journal}{Journal de Physique} \textbf{\bibinfo{volume}{46}},
  \bibinfo{pages}{1345} (\bibinfo{year}{1985}).

\bibitem[{\citenamefont{Claro}(1981)}]{claro1981spectrum}
\bibinfo{author}{\bibfnamefont{F.}~\bibnamefont{Claro}},
  \bibinfo{journal}{phys.\ stat.\ sol.\ (b)} \textbf{\bibinfo{volume}{104}},
  \bibinfo{pages}{K31} (\bibinfo{year}{1981}).

\bibitem[{\citenamefont{Thouless et~al.}(1982)\citenamefont{Thouless, Kohmoto,
  Nightingale, and Den~Nijs}}]{thouless1982quantized}
\bibinfo{author}{\bibfnamefont{D.}~\bibnamefont{Thouless}},
  \bibinfo{author}{\bibfnamefont{M.}~\bibnamefont{Kohmoto}},
  \bibinfo{author}{\bibfnamefont{M.}~\bibnamefont{Nightingale}},
  \bibnamefont{and} \bibinfo{author}{\bibfnamefont{M.}~\bibnamefont{Den~Nijs}},
  \bibinfo{journal}{Phys.\ Rev.\ Lett.} \textbf{\bibinfo{volume}{49}},
  \bibinfo{pages}{405} (\bibinfo{year}{1982}).

\bibitem[{\citenamefont{Novoselov et~al.}(2005)\citenamefont{Novoselov, Geim,
  Morozov, Jiang, Katsnelson, Grigorieva, Dubonos, and
  Firsov}}]{novoselov2005two}
\bibinfo{author}{\bibfnamefont{K.}~\bibnamefont{Novoselov}},
  \bibinfo{author}{\bibfnamefont{A.~K.} \bibnamefont{Geim}},
  \bibinfo{author}{\bibfnamefont{S.}~\bibnamefont{Morozov}},
  \bibinfo{author}{\bibfnamefont{D.}~\bibnamefont{Jiang}},
  \bibinfo{author}{\bibfnamefont{M.}~\bibnamefont{Katsnelson}},
  \bibinfo{author}{\bibfnamefont{I.}~\bibnamefont{Grigorieva}},
  \bibinfo{author}{\bibfnamefont{S.}~\bibnamefont{Dubonos}}, \bibnamefont{and}
  \bibinfo{author}{\bibfnamefont{A.}~\bibnamefont{Firsov}},
  \bibinfo{journal}{Nature} \textbf{\bibinfo{volume}{438}},
  \bibinfo{pages}{197} (\bibinfo{year}{2005}).

\bibitem[{\citenamefont{Hatsugai et~al.}(2006)\citenamefont{Hatsugai, Fukui,
  and Aoki}}]{hatsugai2006topological}
\bibinfo{author}{\bibfnamefont{Y.}~\bibnamefont{Hatsugai}},
  \bibinfo{author}{\bibfnamefont{T.}~\bibnamefont{Fukui}}, \bibnamefont{and}
  \bibinfo{author}{\bibfnamefont{H.}~\bibnamefont{Aoki}},
  \bibinfo{journal}{Phys.\ Rev.\ B} \textbf{\bibinfo{volume}{74}},
  \bibinfo{pages}{205414} (\bibinfo{year}{2006}).

\bibitem[{\citenamefont{Sheng et~al.}(2006)\citenamefont{Sheng, Sheng, and
  Weng}}]{sheng2006quantum}
\bibinfo{author}{\bibfnamefont{D.}~\bibnamefont{Sheng}},
  \bibinfo{author}{\bibfnamefont{L.}~\bibnamefont{Sheng}}, \bibnamefont{and}
  \bibinfo{author}{\bibfnamefont{Z.}~\bibnamefont{Weng}},
  \bibinfo{journal}{Phys.\ Rev.\ B} \textbf{\bibinfo{volume}{73}},
  \bibinfo{pages}{233406} (\bibinfo{year}{2006}).

\bibitem[{\citenamefont{Skyrme}(1962)}]{skyrme1962unified}
\bibinfo{author}{\bibfnamefont{T.~H.~R.} \bibnamefont{Skyrme}},
  \bibinfo{journal}{Nuclear Physics} \textbf{\bibinfo{volume}{31}},
  \bibinfo{pages}{556} (\bibinfo{year}{1962}).

\bibitem[{\citenamefont{Bogdanov and
  Yablonskii}(1989)}]{bogdanov1989thermodynamically}
\bibinfo{author}{\bibfnamefont{A.}~\bibnamefont{Bogdanov}} \bibnamefont{and}
  \bibinfo{author}{\bibfnamefont{D.}~\bibnamefont{Yablonskii}},
  \bibinfo{journal}{Zh. Eksp. Teor. Fiz} \textbf{\bibinfo{volume}{95}},
  \bibinfo{pages}{182} (\bibinfo{year}{1989}).

\bibitem[{\citenamefont{Bogdanov and
  Hubert}(1994)}]{bogdanov1994thermodynamically}
\bibinfo{author}{\bibfnamefont{A.}~\bibnamefont{Bogdanov}} \bibnamefont{and}
  \bibinfo{author}{\bibfnamefont{A.}~\bibnamefont{Hubert}},
  \bibinfo{journal}{J. Magn.\ Magn.\ Mater.} \textbf{\bibinfo{volume}{138}},
  \bibinfo{pages}{255} (\bibinfo{year}{1994}).

\bibitem[{\citenamefont{R{\"o}{\ss}ler
  et~al.}(2006)\citenamefont{R{\"o}{\ss}ler, Bogdanov, and
  Pfleiderer}}]{rossler2006spontaneous}
\bibinfo{author}{\bibfnamefont{U.}~\bibnamefont{R{\"o}{\ss}ler}},
  \bibinfo{author}{\bibfnamefont{A.}~\bibnamefont{Bogdanov}}, \bibnamefont{and}
  \bibinfo{author}{\bibfnamefont{C.}~\bibnamefont{Pfleiderer}},
  \bibinfo{journal}{Nature} \textbf{\bibinfo{volume}{442}},
  \bibinfo{pages}{797} (\bibinfo{year}{2006}).

\bibitem[{\citenamefont{M{\"u}hlbauer et~al.}(2009)\citenamefont{M{\"u}hlbauer,
  Binz, Jonietz, Pfleiderer, Rosch, Neubauer, Georgii, and
  B{\"o}ni}}]{muhlbauer2009skyrmion}
\bibinfo{author}{\bibfnamefont{S.}~\bibnamefont{M{\"u}hlbauer}},
  \bibinfo{author}{\bibfnamefont{B.}~\bibnamefont{Binz}},
  \bibinfo{author}{\bibfnamefont{F.}~\bibnamefont{Jonietz}},
  \bibinfo{author}{\bibfnamefont{C.}~\bibnamefont{Pfleiderer}},
  \bibinfo{author}{\bibfnamefont{A.}~\bibnamefont{Rosch}},
  \bibinfo{author}{\bibfnamefont{A.}~\bibnamefont{Neubauer}},
  \bibinfo{author}{\bibfnamefont{R.}~\bibnamefont{Georgii}}, \bibnamefont{and}
  \bibinfo{author}{\bibfnamefont{P.}~\bibnamefont{B{\"o}ni}},
  \bibinfo{journal}{Science} \textbf{\bibinfo{volume}{323}},
  \bibinfo{pages}{915} (\bibinfo{year}{2009}).

\bibitem[{\citenamefont{Dzyaloshinsky}(1958)}]{dzyaloshinsky1958thermodynamic}
\bibinfo{author}{\bibfnamefont{I.}~\bibnamefont{Dzyaloshinsky}},
  \bibinfo{journal}{J. Phys.\ Chem.\ Sol.} \textbf{\bibinfo{volume}{4}},
  \bibinfo{pages}{241} (\bibinfo{year}{1958}).

\bibitem[{\citenamefont{Moriya}(1960)}]{moriya1960anisotropic}
\bibinfo{author}{\bibfnamefont{T.}~\bibnamefont{Moriya}},
  \bibinfo{journal}{Phys.\ Rev.} \textbf{\bibinfo{volume}{120}},
  \bibinfo{pages}{91} (\bibinfo{year}{1960}).

\bibitem[{\citenamefont{Neubauer et~al.}(2009)\citenamefont{Neubauer,
  Pfleiderer, Binz, Rosch, Ritz, Niklowitz, and
  B{\"o}ni}}]{neubauer2009topological}
\bibinfo{author}{\bibfnamefont{A.}~\bibnamefont{Neubauer}},
  \bibinfo{author}{\bibfnamefont{C.}~\bibnamefont{Pfleiderer}},
  \bibinfo{author}{\bibfnamefont{B.}~\bibnamefont{Binz}},
  \bibinfo{author}{\bibfnamefont{A.}~\bibnamefont{Rosch}},
  \bibinfo{author}{\bibfnamefont{R.}~\bibnamefont{Ritz}},
  \bibinfo{author}{\bibfnamefont{P.}~\bibnamefont{Niklowitz}},
  \bibnamefont{and} \bibinfo{author}{\bibfnamefont{P.}~\bibnamefont{B{\"o}ni}},
  \bibinfo{journal}{Phys.\ Rev.\ Lett.} \textbf{\bibinfo{volume}{102}},
  \bibinfo{pages}{186602} (\bibinfo{year}{2009}).

\bibitem[{\citenamefont{Schulz et~al.}(2012)\citenamefont{Schulz, Ritz, Bauer,
  Halder, Wagner, Franz, Pfleiderer, Everschor, Garst, and
  Rosch}}]{schulz2012emergent}
\bibinfo{author}{\bibfnamefont{T.}~\bibnamefont{Schulz}},
  \bibinfo{author}{\bibfnamefont{R.}~\bibnamefont{Ritz}},
  \bibinfo{author}{\bibfnamefont{A.}~\bibnamefont{Bauer}},
  \bibinfo{author}{\bibfnamefont{M.}~\bibnamefont{Halder}},
  \bibinfo{author}{\bibfnamefont{M.}~\bibnamefont{Wagner}},
  \bibinfo{author}{\bibfnamefont{C.}~\bibnamefont{Franz}},
  \bibinfo{author}{\bibfnamefont{C.}~\bibnamefont{Pfleiderer}},
  \bibinfo{author}{\bibfnamefont{K.}~\bibnamefont{Everschor}},
  \bibinfo{author}{\bibfnamefont{M.}~\bibnamefont{Garst}}, \bibnamefont{and}
  \bibinfo{author}{\bibfnamefont{A.}~\bibnamefont{Rosch}},
  \bibinfo{journal}{Nature Phys.} \textbf{\bibinfo{volume}{8}},
  \bibinfo{pages}{301} (\bibinfo{year}{2012}).

\bibitem[{\citenamefont{Kanazawa et~al.}(2011)\citenamefont{Kanazawa, Onose,
  Arima, Okuyama, Ohoyama, Wakimoto, Kakurai, Ishiwata, and
  Tokura}}]{kanazawa2011large}
\bibinfo{author}{\bibfnamefont{N.}~\bibnamefont{Kanazawa}},
  \bibinfo{author}{\bibfnamefont{Y.}~\bibnamefont{Onose}},
  \bibinfo{author}{\bibfnamefont{T.}~\bibnamefont{Arima}},
  \bibinfo{author}{\bibfnamefont{D.}~\bibnamefont{Okuyama}},
  \bibinfo{author}{\bibfnamefont{K.}~\bibnamefont{Ohoyama}},
  \bibinfo{author}{\bibfnamefont{S.}~\bibnamefont{Wakimoto}},
  \bibinfo{author}{\bibfnamefont{K.}~\bibnamefont{Kakurai}},
  \bibinfo{author}{\bibfnamefont{S.}~\bibnamefont{Ishiwata}}, \bibnamefont{and}
  \bibinfo{author}{\bibfnamefont{Y.}~\bibnamefont{Tokura}},
  \bibinfo{journal}{Phys.\ Rev.\ Lett.} \textbf{\bibinfo{volume}{106}},
  \bibinfo{pages}{156603} (\bibinfo{year}{2011}).

\bibitem[{\citenamefont{Lee et~al.}(2009)\citenamefont{Lee, Kang, Onose,
  Tokura, and Ong}}]{lee2009unusual}
\bibinfo{author}{\bibfnamefont{M.}~\bibnamefont{Lee}},
  \bibinfo{author}{\bibfnamefont{W.}~\bibnamefont{Kang}},
  \bibinfo{author}{\bibfnamefont{Y.}~\bibnamefont{Onose}},
  \bibinfo{author}{\bibfnamefont{Y.}~\bibnamefont{Tokura}}, \bibnamefont{and}
  \bibinfo{author}{\bibfnamefont{N.}~\bibnamefont{Ong}},
  \bibinfo{journal}{Phys.\ Rev.\ Lett.} \textbf{\bibinfo{volume}{102}},
  \bibinfo{pages}{186601} (\bibinfo{year}{2009}).

\bibitem[{\citenamefont{Li et~al.}(2013)\citenamefont{Li, Kanazawa, Yu,
  Tsukazaki, Kawasaki, Ichikawa, Jin, Kagawa, and Tokura}}]{li2013robust}
\bibinfo{author}{\bibfnamefont{Y.}~\bibnamefont{Li}},
  \bibinfo{author}{\bibfnamefont{N.}~\bibnamefont{Kanazawa}},
  \bibinfo{author}{\bibfnamefont{X.}~\bibnamefont{Yu}},
  \bibinfo{author}{\bibfnamefont{A.}~\bibnamefont{Tsukazaki}},
  \bibinfo{author}{\bibfnamefont{M.}~\bibnamefont{Kawasaki}},
  \bibinfo{author}{\bibfnamefont{M.}~\bibnamefont{Ichikawa}},
  \bibinfo{author}{\bibfnamefont{X.}~\bibnamefont{Jin}},
  \bibinfo{author}{\bibfnamefont{F.}~\bibnamefont{Kagawa}}, \bibnamefont{and}
  \bibinfo{author}{\bibfnamefont{Y.}~\bibnamefont{Tokura}},
  \bibinfo{journal}{Phys.\ Rev.\ Lett.} \textbf{\bibinfo{volume}{110}},
  \bibinfo{pages}{117202} (\bibinfo{year}{2013}).

\bibitem[{\citenamefont{Hamamoto et~al.}(2015)\citenamefont{Hamamoto, Ezawa,
  and Nagaosa}}]{hamamoto2015quantized}
\bibinfo{author}{\bibfnamefont{K.}~\bibnamefont{Hamamoto}},
  \bibinfo{author}{\bibfnamefont{M.}~\bibnamefont{Ezawa}}, \bibnamefont{and}
  \bibinfo{author}{\bibfnamefont{N.}~\bibnamefont{Nagaosa}},
  \bibinfo{journal}{Phys.\ Rev.\ B} \textbf{\bibinfo{volume}{92}},
  \bibinfo{pages}{115417} (\bibinfo{year}{2015}).

\bibitem[{\citenamefont{Lado and
  Fern{\'a}ndez-Rossier}(2015)}]{lado2015quantum}
\bibinfo{author}{\bibfnamefont{J.~L.} \bibnamefont{Lado}} \bibnamefont{and}
  \bibinfo{author}{\bibfnamefont{J.}~\bibnamefont{Fern{\'a}ndez-Rossier}},
  \bibinfo{journal}{Phys.\ Rev.\ B} \textbf{\bibinfo{volume}{92}},
  \bibinfo{pages}{115433} (\bibinfo{year}{2015}).

\bibitem[{\citenamefont{Okubo et~al.}(2012)\citenamefont{Okubo, Chung, and
  Kawamura}}]{okubo2012multiple}
\bibinfo{author}{\bibfnamefont{T.}~\bibnamefont{Okubo}},
  \bibinfo{author}{\bibfnamefont{S.}~\bibnamefont{Chung}}, \bibnamefont{and}
  \bibinfo{author}{\bibfnamefont{H.}~\bibnamefont{Kawamura}},
  \bibinfo{journal}{Phys.\ Rev.\ Lett.} \textbf{\bibinfo{volume}{108}},
  \bibinfo{pages}{017206} (\bibinfo{year}{2012}).

\bibitem[{\citenamefont{Nagaosa et~al.}(2010)\citenamefont{Nagaosa, Sinova,
  Onoda, MacDonald, and Ong}}]{nagaosa2010anomalous}
\bibinfo{author}{\bibfnamefont{N.}~\bibnamefont{Nagaosa}},
  \bibinfo{author}{\bibfnamefont{J.}~\bibnamefont{Sinova}},
  \bibinfo{author}{\bibfnamefont{S.}~\bibnamefont{Onoda}},
  \bibinfo{author}{\bibfnamefont{A.}~\bibnamefont{MacDonald}},
  \bibnamefont{and} \bibinfo{author}{\bibfnamefont{N.}~\bibnamefont{Ong}},
  \bibinfo{journal}{Rev.\ Mod.\ Phys.} \textbf{\bibinfo{volume}{82}},
  \bibinfo{pages}{1539} (\bibinfo{year}{2010}).

\bibitem[{\citenamefont{Hatsugai}(1993{\natexlab{a}})}]{Hatsugai1993}
\bibinfo{author}{\bibfnamefont{Y.}~\bibnamefont{Hatsugai}},
  \bibinfo{journal}{Phys.\ Rev.\ B} \textbf{\bibinfo{volume}{48}},
  \bibinfo{pages}{11851} (\bibinfo{year}{1993}{\natexlab{a}}).

\bibitem[{\citenamefont{Hatsugai}(1993{\natexlab{b}})}]{Hatsugai1993a}
\bibinfo{author}{\bibfnamefont{Y.}~\bibnamefont{Hatsugai}},
  \bibinfo{journal}{Phys.\ Rev.\ Lett.} \textbf{\bibinfo{volume}{71}},
  \bibinfo{pages}{3697–3700} (\bibinfo{year}{1993}{\natexlab{b}}).

\bibitem[{\citenamefont{Everschor-Sitte and Sitte}(2014)}]{everschor2014real}
\bibinfo{author}{\bibfnamefont{K.}~\bibnamefont{Everschor-Sitte}}
  \bibnamefont{and} \bibinfo{author}{\bibfnamefont{M.}~\bibnamefont{Sitte}},
  \bibinfo{journal}{J. Appl.\ Phys.} \textbf{\bibinfo{volume}{115}},
  \bibinfo{pages}{172602} (\bibinfo{year}{2014}).

\bibitem[{\citenamefont{Ohgushi et~al.}(2000)\citenamefont{Ohgushi, Murakami,
  and Nagaosa}}]{ohgushi2000spin}
\bibinfo{author}{\bibfnamefont{K.}~\bibnamefont{Ohgushi}},
  \bibinfo{author}{\bibfnamefont{S.}~\bibnamefont{Murakami}}, \bibnamefont{and}
  \bibinfo{author}{\bibfnamefont{N.}~\bibnamefont{Nagaosa}},
  \bibinfo{journal}{Phys.\ Rev.\ B} \textbf{\bibinfo{volume}{62}},
  \bibinfo{pages}{6065(R)} (\bibinfo{year}{2000}).

\bibitem[{\citenamefont{Hsu and Falicov}(1976)}]{hsu1976level}
\bibinfo{author}{\bibfnamefont{W.~Y.} \bibnamefont{Hsu}} \bibnamefont{and}
  \bibinfo{author}{\bibfnamefont{L.~M.} \bibnamefont{Falicov}},
  \bibinfo{journal}{Phys.\ Rev.\ B} \textbf{\bibinfo{volume}{13}},
  \bibinfo{pages}{1595} (\bibinfo{year}{1976}).

\bibitem[{\citenamefont{Arai and Hatsugai}(2009)}]{arai2009quantum}
\bibinfo{author}{\bibfnamefont{M.}~\bibnamefont{Arai}} \bibnamefont{and}
  \bibinfo{author}{\bibfnamefont{Y.}~\bibnamefont{Hatsugai}},
  \bibinfo{journal}{Phys.\ Rev.\ B} \textbf{\bibinfo{volume}{79}},
  \bibinfo{pages}{075429} (\bibinfo{year}{2009}).

\bibitem[{\citenamefont{Lifshitz et~al.}(1957)\citenamefont{Lifshitz, Azbel,
  and Kaganov}}]{lifshitz1957theory}
\bibinfo{author}{\bibfnamefont{I.}~\bibnamefont{Lifshitz}},
  \bibinfo{author}{\bibfnamefont{M.~I.} \bibnamefont{Azbel}}, \bibnamefont{and}
  \bibinfo{author}{\bibfnamefont{M.}~\bibnamefont{Kaganov}},
  \bibinfo{journal}{Sov.\ Phys.\ JETP} \textbf{\bibinfo{volume}{4}},
  \bibinfo{pages}{41} (\bibinfo{year}{1957}).

\bibitem[{\citenamefont{Arai and Hatsugai}(2010)}]{arai2010numerical}
\bibinfo{author}{\bibfnamefont{M.}~\bibnamefont{Arai}} \bibnamefont{and}
  \bibinfo{author}{\bibfnamefont{Y.}~\bibnamefont{Hatsugai}},
  \bibinfo{journal}{Phys.\ E} \textbf{\bibinfo{volume}{42}},
  \bibinfo{pages}{740} (\bibinfo{year}{2010}).

\bibitem[{\citenamefont{Yu et~al.}(2010)\citenamefont{Yu, Onose, Kanazawa,
  Park, Han, Matsui, Nagaosa, and Tokura}}]{yu2010real}
\bibinfo{author}{\bibfnamefont{X.}~\bibnamefont{Yu}},
  \bibinfo{author}{\bibfnamefont{Y.}~\bibnamefont{Onose}},
  \bibinfo{author}{\bibfnamefont{N.}~\bibnamefont{Kanazawa}},
  \bibinfo{author}{\bibfnamefont{J.}~\bibnamefont{Park}},
  \bibinfo{author}{\bibfnamefont{J.}~\bibnamefont{Han}},
  \bibinfo{author}{\bibfnamefont{Y.}~\bibnamefont{Matsui}},
  \bibinfo{author}{\bibfnamefont{N.}~\bibnamefont{Nagaosa}}, \bibnamefont{and}
  \bibinfo{author}{\bibfnamefont{Y.}~\bibnamefont{Tokura}},
  \bibinfo{journal}{Nature} \textbf{\bibinfo{volume}{465}},
  \bibinfo{pages}{901} (\bibinfo{year}{2010}).

\bibitem[{\citenamefont{Yu et~al.}(2011)\citenamefont{Yu, Kanazawa, Onose,
  Kimoto, Zhang, Ishiwata, Matsui, and Tokura}}]{yu2011near}
\bibinfo{author}{\bibfnamefont{X.}~\bibnamefont{Yu}},
  \bibinfo{author}{\bibfnamefont{N.}~\bibnamefont{Kanazawa}},
  \bibinfo{author}{\bibfnamefont{Y.}~\bibnamefont{Onose}},
  \bibinfo{author}{\bibfnamefont{K.}~\bibnamefont{Kimoto}},
  \bibinfo{author}{\bibfnamefont{W.}~\bibnamefont{Zhang}},
  \bibinfo{author}{\bibfnamefont{S.}~\bibnamefont{Ishiwata}},
  \bibinfo{author}{\bibfnamefont{Y.}~\bibnamefont{Matsui}}, \bibnamefont{and}
  \bibinfo{author}{\bibfnamefont{Y.}~\bibnamefont{Tokura}},
  \bibinfo{journal}{Nature Mater.} \textbf{\bibinfo{volume}{10}},
  \bibinfo{pages}{106} (\bibinfo{year}{2011}).

\bibitem[{\citenamefont{Matsuno et~al.}(2016)\citenamefont{Matsuno, Ogawa,
  Yasuda, Kagawa, Koshibae, Nagaosa, Tokura, and
  Kawasaki}}]{matsuno2016interface}
\bibinfo{author}{\bibfnamefont{J.}~\bibnamefont{Matsuno}},
  \bibinfo{author}{\bibfnamefont{N.}~\bibnamefont{Ogawa}},
  \bibinfo{author}{\bibfnamefont{K.}~\bibnamefont{Yasuda}},
  \bibinfo{author}{\bibfnamefont{F.}~\bibnamefont{Kagawa}},
  \bibinfo{author}{\bibfnamefont{W.}~\bibnamefont{Koshibae}},
  \bibinfo{author}{\bibfnamefont{N.}~\bibnamefont{Nagaosa}},
  \bibinfo{author}{\bibfnamefont{Y.}~\bibnamefont{Tokura}}, \bibnamefont{and}
  \bibinfo{author}{\bibfnamefont{M.}~\bibnamefont{Kawasaki}},
  \bibinfo{journal}{Science Advances} \textbf{\bibinfo{volume}{2}},
  \bibinfo{pages}{e1600304} (\bibinfo{year}{2016}).

\end{thebibliography}
\bibliographystyle{apsrev}

\end{document}